\documentclass[12pt]{aastex}
\usepackage{emulateapj5}
\shorttitle{Development of thermal instability}
\begin{document}

\def\VS{V\'azquez-Semadeni}
\def\x{{\bf x}}

\title{The Temperature Distribution in Turbulent Interstellar Gas}
\author{A. Gazol}
\affil{Instituto de Astronom\'\i a, UNAM, Apdo. Postal
3-72, (Xangari) Morelia, Michoac\'an, 58089, MEXICO}
\email{a.gazol@astrosmo.unam.mx}

\author{E. V\'azquez-Semadeni}
\affil{Instituto de Astronom\'\i a, UNAM, Apdo. Postal
3-72, (Xangari) Morelia, Michoac\'an, 58089, MEXICO}
\email{e.vazquez@astrosmo.unam.mx}

\author{F. J. S\'anchez-Salcedo}
\affil{Instituto de Astronom\'\i a, UNAM, Apdo. Postal 70-264,
M\'exico D.F., 04510, MEXICO}
\email{jsanchez@astroscu.unam.mx}

\and 

\author{J. Scalo}
\affil{Astronomy Department, University of Texas, Austin, TX 78712}
\email{parrot@astro.as.utexas.edu}

%

\begin{abstract}
We discuss the temperature distribution in a two-dimensional,
thermally unstable numerical simulation of the warm and cold gas in the
Galactic disk, including the magnetic field, self-gravity, the
Coriolis force, stellar energy injection and a realistic cooling function.
We find that $\sim 50\%$ of the turbulent gas mass has temperatures
in what would be the thermally unstable range if thermal instability were to 
be considered alone. This appears to be a consequence of there being
many other forces at play than just thermal pressure. 
We also point out that a bimodal temperature pdf is a simple 
consequence of the form of the interstellar cooling function and is
not necessarily a signature of discontinuous phase transitions.
\end{abstract}

\section{Introduction}

The isobaric mode of the thermal instability (hereafter TI; Field 1965)
is the basis for the ``two-phase" model of Field, Goldsmith, \& Habing
(1969) for the interstellar medium (ISM), aimed at
explaining the existence of diffuse interstellar clouds. 
The development of this mode of TI produces a phase 
segregation leading to dense clouds in pressure equilibrium 
with their surroundings. This picture is still fundamental in the classical
three-phase model of the ISM advanced by McKee \& Ostriker (1977,
hereafter MO), in which 
supernova explosions produce expanding bubbles of hot gas that sweep
up the ambient medium, collecting it into shells that cool and fragment
into clouds of cold ($T\lesssim 10^2$K) gas. In this model, the warm
medium (neutral and ionized, $T\lesssim 10^4$K) forms at the 
interfaces between the hot and cold gas, due to the soft x-ray radiation
from the hot gas partially penetrating the clouds. Thus, the swept-up
gas consists of two distinct neutral phases in pressure equilibrium,
each in thermally stable equilibrium, with the hot phase also in pressure
equilibrium, although the model allows for a distribution of pressures
in the ISM.

The MO model has been questioned on a number of
counts (e.g., Cox 1995; Elmegreen 1997). In particular, 
in spite of having been constructed around the idea of a ``violent'' ISM
dominated by supernovae (McCray \& Snow 1979), the model focused
mainly on radiative and evaporative processes, under the 
fundamental premise of thermal pressure equilibrium, but 
neglected other dynamical agents, such as self-gravity and the Coriolis
and Lorentz forces. Elmegreen (1991, 1994) presented an instability
analysis including many of those extra agents, finding that the
instability is of a very different nature than TI alone. 
Moreover, the ISM, being continually
stirred by the energy injection from massive stars and possibly other
sources as well, is expected to be highly 
turbulent (see, e.g., the reviews by Scalo 1987 and V\'azquez-Semadeni
et al.\ 2000b, and the volume by Franco \& Carrami\~nana 1999). 
In a previous paper (V\'azquez-Semadeni, Gazol \& Scalo 2000a, hereafter
Paper I), we presented low resolution ($128^2$) numerical simulations of
the warm and cold interstellar gas on a maximun scale of $1$ kpc. These
thermally unstable simulations, which included magnetic fields, the Coriolis 
force, self-gravity, and turbulence driven by stellar-like sources
(but not supernova explosions) located
at the (negative dilatation) density maxima, showed that the density 
probability distribution
function (PDF) is substantially 
different from the bimodal one  expected for a medium subject to the isobaric 
mode of TI alone.  Apparently, the gas redistribution promoted by 
stellar activity combined with the magnetic pressure and Coriolis force
erased the signature of TI in the density PDF. 

An important related question concerns the PDF of the gas
temperature. Observationally, the latter is studied primarily 
through the HI 21-cm hyperfine line. 
Dickey, Salpeter, \& Terzian (1977) used histograms of HI absorption
line spin temperatures along high-latitude lines of sight to claim
that a significant fraction of the neutral HI has temperatures in a
range that is inconsistent with the two-phase model of the ISM.
More recent work has vacillated on this point.
For the warm neutral medium, a tentative 
lower limit of $5000$ K was proposed by Kulkarni \& Heiles (1987) using
the available data at that time. However, interferometric 
(Kalberla, Schwartz, \& Goss 1985) and optical/UV absorption-line
(Spitzer \& Fitzpatrick 1995; Fitzpatrick \& Spitzer 1997) measurements
indicate the presence of warm neutral hydrogen with lower temperatures, 
which lie in the thermally unstable range. More recently,
Heiles (2001) has presented new observations suggesting that actually a
substantial fraction or even the majority of the gas 
in the warm neutral medium is at thermally unstable temperatures. 
Heiles' temperatures are based on linewidths, and hence could be
overestimated if part of the linewidth is non-thermal. In any case, they 
cannot be lower than 500 K, as they are not seen in absorption (Heiles 2001).
These observations support the initial contention of Dickey et al. (1977), and
are in clear disagreement with 2- or 3-phase models, 
in which no gas at unstable temperatures is expected. 

 From the theoretical point of view, several models that predict gas in
the thermally unstable region have been proposed.
A time-dependent but non-hydrodynamic
model for the ISM in which random supernova x-ray flashes
occasionally heat the otherwise cooling gas was examined in detail by
Gerola et al. (1974). Dalgarno \& McCray (1972) showed  how the
temperature pdf is simply and in general related to the shape of the
cooling function and the rate of stochastic flashes, and estimated the
fraction of material in the 
unstable temperature range for two different supernova rates.  Lioure \&
Chi\'eze (1990) obtained a similar result by assuming a constant mass flux  
among a number of idealized ``phases", assuming isobaric, constant-heating,
and  non-hydrodynamic evolution.  The
``scale-dependent phase continuum" model of the ISM by Norman \&
Ferrara (1996) also contains gas in the thermally unstable range,
because of a combination of spatially-averaged stellar and turbulent
heating. 
Numerically, temperature histograms from 3D MHD simulations of the ISM 
including essentially the same physical ingredients as those in Paper I plus
supernova explosions have been presented by Korpi et al.\ (1999). In
those PDFs, substantial amounts of gas are seen to be at temperatures
inbetween the cold and warm phases of the ISM. Unfortunately, those
authors chose to use cooling functions that only imply a thermally
unstable range between the cold and warm gas when no
heating is present (see Burkert \& Lin 2000). Although no global heating was 
used in those simulations, the presence of stellar heating makes the
role of the isobaric mode of TI unclear.

The aim of this paper is to provide additional, numerical evidence that a
substantial fraction of the mass in the turbulent ISM, even in the
presence of TI, is expected at thermally unstable temperatures, when
other relevant physical agents are considered besides thermal
pressure. To this end, we present  temperature histograms from two-dimensional,
ISM-like, thermally unstable simulations using a more realistic cooling 
function than the ones used in Paper I. 

\section{The numerical model}
We use the numerical model presented by Passot, V\'azquez-Semadeni \& Pouquet 
(1995), which uses a single-fluid approach to describe the interstellar gas 
on the galactic disk at the solar Galactocentric distance.

The fiducial simulation presented here represents 1 kpc$^2$ on the
Galactic plane at the solar Galactocentric distance, and 
includes self-gravity, the magnetic field,
and disk rotation, as well as model terms for the radiative cooling, the
diffuse background radiation and the local thermal energy input due to
star formation (SF).
The equations describing the evolution are solved in two dimensions, at a 
resolution of $512^2$ grid points, by means of a pseudo-spectral method, 
which imposes periodic boundary conditions. Since this method produces
no numerical viscosity of diffusivity, diffusive terms, with their
respective coefficients, are included in the equations (see Passot et
al.\ 1995). This allows us to perform convergence tests by simply
varying these coefficients.

Stellar-, ionization-like heating is applied at small scales ($\sim 7$
pixels FWHM) and consists of local, 
discrete heating sources turned on  at a grid point $x$ whenever  
$\rho(\x)>\rho_{\rm SF}$ and $\nabla\cdot u(\x)<0$, where $\rho_{\rm SF}$ 
is a free parameter, but taken equal to $15\langle\rho\rangle$, where
$\langle\rho\rangle=1$ cm$^{-3}$. 
The sources stay on for a time $\Delta t=6\times
10^6$yr. For further details, see Passot et al.\ (1995).
 
The background heating is taken as a constant 
($\Gamma_0=2.51 \times 10^{-26} {\rm erg}\;{\rm s}^{-1} {\rm H}^{-1}$,
where $H^{-1}$ means ``per Hydrogen atom''). 
We use its value to fit the ``standard'' equilibrium $P$ vs. $\rho$ curve of 
Wolfire et al. (1995) assuming that the background heating is in equilibrium 
with a cooling function that has a piece-wise power-law dependence on the 
temperature with the form
\begin{equation}
\Lambda=C_{i,i+1} T^{\beta_{i,i+1}}  \;\;\;\; {\rm for} \;\;\; T_{i}\leq
T< T_{i+1}. 
\end{equation}
The resulting values of the coefficients, exponents and transition temperatures
are given in S\'anchez-Slacedo, V\'azquez-Semadeni \& Gazol
(2001). Under thermal equilibrium
conditions, the gas is thermally unstable in the isobaric mode for 
$313\;{\rm K}< T <6102\;{\rm K}$  ($\beta_{34}=0.56$), and marginally stable 
for $141\;{\rm K}< T <313\;{\rm K}$ ($\beta_{23}=1.0$). In thermal
equilibrium, the ``boundary'' temperatures $T_{4,3,2}=6102$, 313 and 141 K
correspond to densities $\rho_{4,3,2} = 0.60$, 3.2 and 7.1 cm$^{-3}$,
respectively. These are all 
below $\rho_{\rm SF}$, but the latter in turn is smaller than the
density that would correspond to purely isobaric evolution from the
initial conditions, $\rho_P\approx 50$ cm$^{-3}$ (S\'anchez-Salcedo et
al.\ 2001). We do not expect this to be a problem, however, since for
our purposes it is sufficient that the stellar activity does not impede
the formation of the dense stable phase. Moreover, as will be
seen below, most of the gas in the unstable regime is not directly
produced by stellar activity anyway.

The initial random fluctuations are Gaussian with random phases, with a
spectrum that peaks at $k=4$, where $k$ is the wavenumber in units of
the inverse box length. The initial velocity, density and temperature
fluctuations have an rms amplitude of 0.3, while those of the magnetic
field have an amplitude of 1, all in normalized code units ($\rho_0 =1$
cm$^{-3}$, $T_0=10^4$ K, $u_0=11.7$ km s$^{-1}$, $B_0=5\,\mu{\rm G}$). 
The magnetic field has also an initial azimuthal uniform component of
strength $1.5\mu{\rm G}$. A rotation rate $\Omega=2 \pi/(2 \times
10^8$yr) around the Galactic center is imposed, as is a sinusoidal shear 
rate (see Passot et al.\ 1995) of amplitude 8.8 km s$^{-1}$ kpc$^{-1}$.

These initial conditions and the realistic cooling function and
parameters we use are such that the
system is not only unstable thermally, but also with respect to
the combined instability criterion of  
Elmegreen 1994 without shear (see Passot et al.\ 1995 and Paper I for the 2D
case). Our simulations are unstable even in the
presence of shear, as turning off SF causes them to produce condensations of
densities up to $\rho \sim 200$ cm$^{-3}$, and eventually stop because
of the steep gradients produced.

\section{Results and discussion} \label{sec:discussion}

\begin{figure}
\plotone{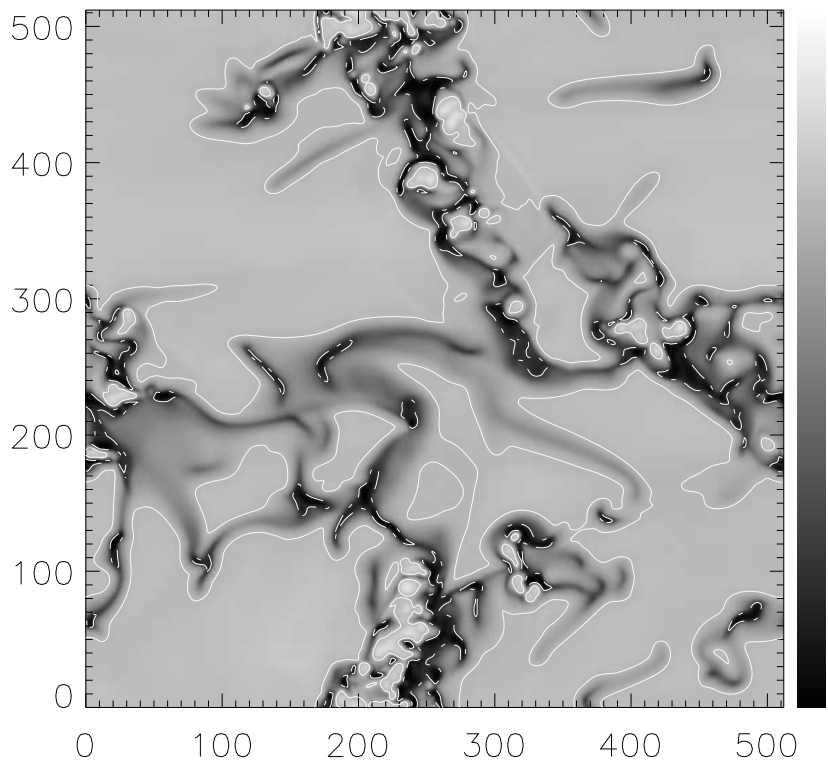}
\caption{Temperature field at $t=5.03\times 10^7$ yr. The gray scale is
logarithmic with darker tones at cooler regions. The maximum and minimum
value are $1.77\times 10^4$ K and $95$ K, respectively. 
Superimposed contours show the upper ({\it solid line}) and lower ({\it
dashed line}) limits of the unstable range. Note that much of the gas in 
the unstable regime is not directly associated with stellar activity.}
\label{fig:temp}
\end{figure}

\begin{figure}
\plotone{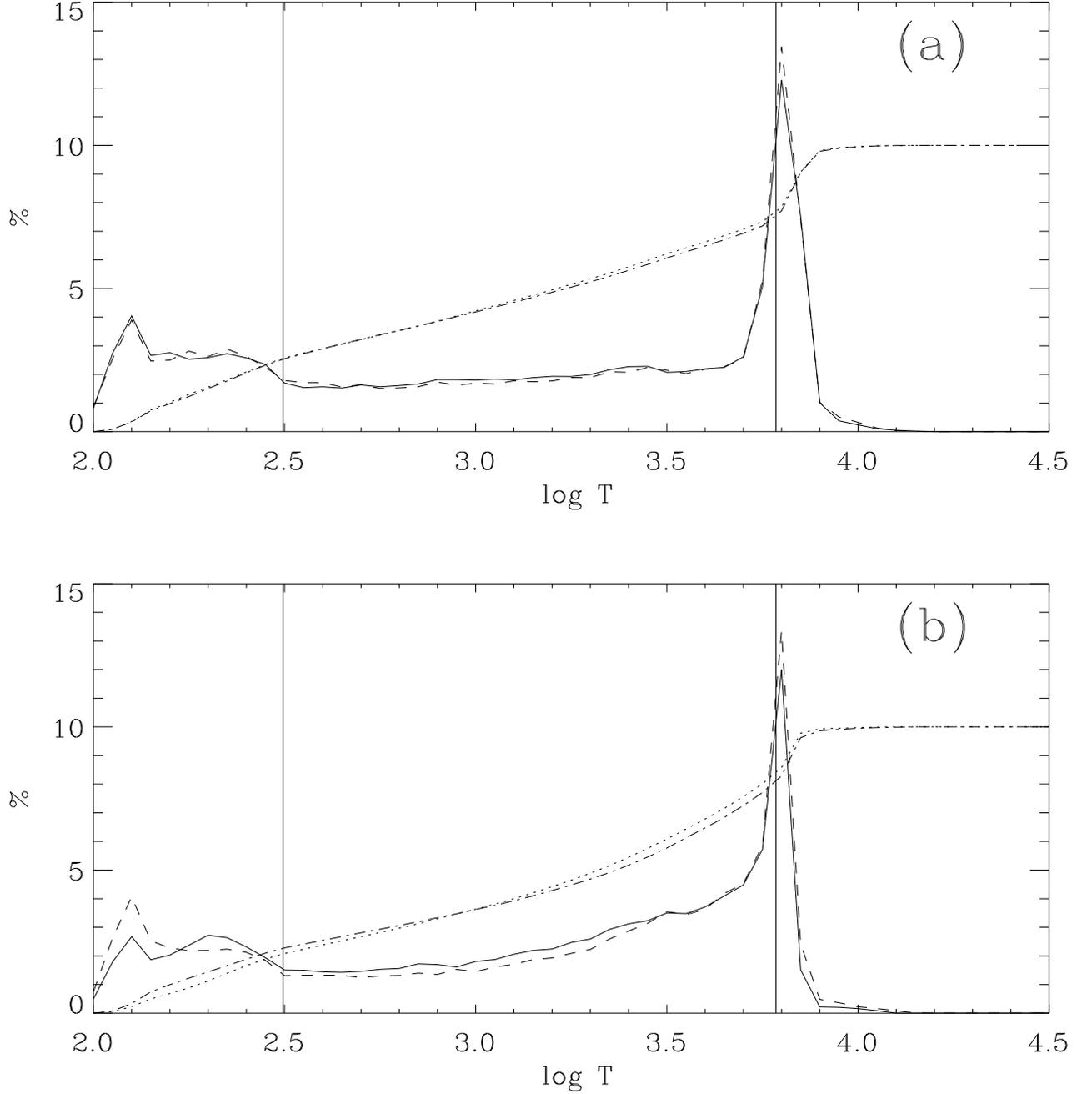}
\caption{Density-weighted temperature histograms and cumulative distributions
(divided by 10). In both panels, {\it solid} and {\it dotted} lines
refer to the fiducial simulation shown in fig. \ref{fig:temp}, and {\it
dashed and dashed-dotted lines} refer to comparison runs for testing
convergence : {\it Top} (a), comparison with a run with 
thermal diffusion decreased by a factor of 5 at $t=5.03\times 10^7$
yr. {\it Bottom} (b), comparison with a run with all diffusivities
decreased by a factor of $\sim 2$ at $t=2.04\times 10^7$ yr. In the top
panel, in which a stationary state has been reached, the cumulative
distribution shows that $\sim 50\%$ of the mass is 
in the unstable temperature range.}
\label{fig:hist}
\end{figure}

The fiducial simulation reaches a stationary regime after $\sim
3.40\times 10^7$ yr. Figure \ref{fig:temp} shows an image of the
temperature field at $t=5.03 \times 10^7$ yr. The contours show the
upper ({\it solid line}) and lower ({\it dashed line}) boundary
temperatures of the unstable range, $T_4$ and $T_3$. The small bright
spots partially surrounded by 
cool dense gas are the stellar heating sites. For this field, fig.\
\ref{fig:hist}a shows the logarithmic, density-weighted temperature 
histogram ({\it solid line}) and the cumulative distribution ({\it
dotted line}). The vertical lines indicate $T_4$ and $T_3$. 
Although the histogram is
clearly bimodal, with peaks at temperatures just outside the unstable
range, a roughly constant mass fraction per logarithmic temperature
interval is seen to occur in the unstable regime. Given the large
extension of the unstable range, the total mass in this range
is $\sim 50$\% of the mass in the simulation, with $\sim 25$\% in both
the cold and warm stable regimes. We have verified that the presence
of the warm unstable gas is not due to diffusion. The temperature histogram 
does not change appreciably by decreasing the thermal diffusion term by a 
factor of 5 (fig.\ \ref{fig:hist}a; 
{\it dashed} and {\it dashed-dotted lines}). We also
performed a simulation with all diffusivities reduced by factors $\sim 2$, 
and, although it cannot go beyond $t=3.26 \times 10^7$ yr, the cumulative 
distribution of this simulation agrees to within a few percent with that 
of our fiducial run at that time throughout the temperature range 
(fig.\ \ref{fig:hist}b). Finally, we have also checked
that the temperature PDF at a later time ($t = 6.53 \times 10^7$ yr) is
virtually identical to that shown in fig.\ \ref{fig:hist}a,
reassuring us that a statistically stationary state has been reached.

In order to understand these results, it is helpful to briefly describe the
evolution of the simulation. As in Paper I, the initial density and velocity
fluctuations generate filaments (recall the simulation is 2D), which
can later fragment, redisperse, get stretched by the shear, collide with
other filaments to either merge or disrupt, or any
combination thereof. Most of
the initial filaments do not form stars by themselves, except for those
formed by the strongest initial fluctuations, which evolve in a manner
similar to that described by Henebelle \& P\'erault (1999). Some filaments
redisperse, apparently under the action of the other forces at play in
the simulation, such as shear and tidal stresses, and magnetic
pressure. 

In most cases, the first SF events occur upon the collision of
filaments, but subsequently SF auto-propagates, with new stars forming
in the shells created by previous ones, although this self-propagation
proceeds slowly due to the neglect of supernovae. The stellar heating in the
simulation produces bubbles of warm gas that in general reach rather
high temperatures (up to $\sim$ 15,000 K) while the stars are
on. Individual stars hardly affect their parent cloud, but stars formed
in ``associations'' create bubbles that merge with the warm, diffuse
medium. After the stars turn off, these regions are left out of thermal
equilibrium and, since they have also been evacuated to rather low
densities ($\sim 0.2$--1 cm$^{-3}$), they require times $\sim 2$ Myr to
return to thermal equilibrium {\it in the warm ``stable'' range}. At
advanced evolutionary stages, 
the general impression is much more
that of a temperature continuum than one of a two-phase medium. 
Regions in the thermally ``unstable'' range do not exhibit any systematic
tendency to be systematically destroyed. This is most noticeable from watching
the evolution of the contours at the ``transition''
temperatures.\footnote{An animation of this simulation can be retrieved
from http://www.astrosmo.unam.mx/$\sim$e.vazquez/turbulence\_HP/video/TIrun9.mpg.} The contours bounding
the unstable range in general do not appear to have a tendency to 
approach one another, but instead move in a typically advective fashion.
This is in agreement with the nearly constant mass
fraction ($\sim 50\%$) at unstable temperatures present in the
simulation at various times. In addition, even though star formation
self-propagates, many of the unstable regions in our simulation
have never been under the influence of the stellar heating.
This suggests that the other forces at play
besides thermal pressure provide a net restoring force even in the
unstable regime with respect to TI, in such a way that no abrupt phase
transition occurs. We plan to investigate the
detailed interplay between the various physical agents in a future paper.

It should be emphasized that the bimodal form of the histograms shown in
fig.\ \ref{fig:hist} does not imply that distinct phases must
exist. It is sufficient that, under thermal equilibrium
conditions, the equilibrium temperature $T_{\rm eq}$, plotted as a
function of density, have extended nearly flat portions (``plateaus'')
joined by short density intervals in which $T_{\rm eq}$ varies
rapidly. Thus, even for a smoothly varying density distribution, the
plateau temperatures will in general be more frequent than those at the
intervals of rapid variation, with no need for a sharp (phase)
transition from one temperature to the other. This will give a
temperature PDF similar to those in fig.\ \ref{fig:hist}, with an excess 
at the plateau temperatures, but a non-zero population at intermediate
temperatures.

\section{Conclusions}\label{sec:conclusions}

In this Letter, we have shown that in our 2D simulations including the
isobaric mode of TI, together with the magnetic field, self-gravity and
energy injection mimicking that from OB-star ionizing radiation (but
without supernovae), nearly half of the mass is at thermally unstable
temperatures as shown by density-weighted temperature histograms. This
is apparently partly the result of the other forces at play overwhelming the
``crushing'' drive of TI and restoring a ``normal'' (rather than
reversed) effective pressure 
gradient, in such a way that there is no abrupt transition between the
cold and warm gas, but rather there is a continuous temperature
distribution.  This is a different mechanism for maintaining
the gas in the unstable range than the one in the models of Gerola et
al.\ (1974), in which gas is heated and driven far from equilibrium at
random times by
stellar energy injection, creating a population of gas that
traverses the ``unstable'' range as it cools. In this case, the simplified 
example given by McCray \& Dalgarno (1972) suggests that the fraction
of gas in this temperature range should depend on the star formation
rate. In our simulations this cannot occur, as the gas is subject to
global uniform background heating, so once it reaches the warm stable
phase it has no tendency to cool further. But in the real ISM, in which
the background heating is not uniform, but is strongest in the vicinity
of stellar energy sources, this process is a real additional possibility.

The functional form of the cooling still affects the
temperature PDF, producing peaks at the temperatures that would
be stable under TI alone, but with a substantial fraction of the gas mass
dwelling in the ``unstable'' range at all times. This result strengthens
the view in Paper I that, under the presence of the many other physical
ingredients relevant in the ISM, TI becomes a second-order effect, and
suggests that dynamical processes should not be neglected in
comprehensive models of the ISM.

It is important to note that, because of numerical limitations, our
simulations do not include supernovae, and thus contain no hot gas.
If supernovae were present, the cavities formed by them would
reach temperatures $\sim 10^6$ K and expand much more than the stellar
bubbles do here, due to the
presence of the isochoric mode of TI above $\sim 10^5$ K. However, 
two lines of argument suggest that our results should hold even in the
presence of supernovae. First, if the filling factor of the hot gas at
the Galactic midplane is not too large ($\sim 20\%$; see, e.g.,
Ferri\`ere 1998; Gazol-Pati\~no \& Passot 1999; Avillez 2000), then our
simulation can be thought of as representing the regions of the midplane
not occupied by the hot gas. Second, the simulations of Korpi et al.\
(1999), which do include supernovae, albeit with an uncertain role of
the isobaric mode of TI, give temperature histograms consistent in the
cold and warm ranges with the ones presented in this paper. A definitive test
on  whether the presence of such hot gas would alter our conclusions
will be provided by simulations including supernovae, which we intend to 
present, in 3D and using a different numerical scheme, in a future paper.

\acknowledgements

We are glad to acknowledge interesting discussions with Carl
Heiles. This work has received partial funding from CONACYT grant
27752-E to E.V.-S.

\clearpage

\end{document}